\documentclass[%
 reprint,
 amsmath,amssymb,
 aps,
]{revtex4-1}

\usepackage{graphicx}
\usepackage{dcolumn}
\usepackage{bm}

\allowdisplaybreaks

\begin{document}

\preprint{APS/123-QED}

\title{K-shell ionization and characteristic x-ray radiation by high-energy electrons in multifoil targets}

\author{S.V. Trofymenko}
\email{trofymenko@kipt.kharkov.ua}
\affiliation{Akhiezer Institute for Theoretical Physics of NSC ``Kharkiv Institute of Physics and Technology'', 1 Akademichna st., 61108 Kharkiv, Ukraine}
\affiliation{Karazin Kharkiv National University, 4 Svobody sq., 61022 Kharkiv, Ukraine}


\date{\today}

\begin{abstract}
Processes of K-shell ionization and accompanying characteristic x-ray radiation (CXR) by high-energy electrons moving through a multifoil copper target are considered. Expressions describing the main characteristics of these processes are derived. It is shown that the average K-shell ionization cross section in the target is not defined just by the target material and the electron energy, but also depends on the number of foils in the target, their thickness and separation between them. The corresponding CXR yield is therefore not unambiguously defined by the aggregated target thickness, but depends on the above parameters as well. It is demonstrated that the average K-shell ionization cross section in a multifoil target can be several times larger than the conventional cross section of this process without the density effect impact. This results in a considerable enhancement of CXR yield from the multifoil target compared to the case of electron incidence upon a single foil of the same aggregated thickness. Comparison of CXR yield in the considered case with the yields of some other types of x-ray emission in multifoil targets is made.

\begin{description}
\item[PACS numbers]
\end{description}
\end{abstract}

\pacs{Valid PACS appear here}
\maketitle


\section{Introduction}
\label{Introduction}

Passage of high-energy charged particles through matter is accompanied by ionization of atomic shells, which leads to emission of photons or Auger electrons as a result of recombination of these shells. This process is of special interest for inner atomic shells (particularly, K shells) since the emitted photons in this case belong to the x-ray range and are rather weakly absorbed. Such an emission, known as characteristic x-ray radiation (CXR), is widely applied as a source of monochromatic x-rays and for spectroscopy purposes. 

A series of experimental \cite{Middleman,Dangerfield,Ishii,Hoffmann,Kamiya,Genz,Bak1983,Bak1986,Meyerhof,Spooner} and theoretical \cite{Bak1986,Sorensen1986,Ermilova,Sorensen,Chechin} works has been devoted to study of cross section of K-shell ionization by high-energy particles with the aim of investigating the influence of medium polarization upon it. Such an influence is known as the density effect \cite{Fermi,Sternheimer}, which leads to partial suppression of particle ionization loss at high energies. It was discovered that the cross section is significantly influenced by the transition radiation (TR) emitted upon the particle entrance into the foil. The contribution of such radiation to ionization of K shells leads to complete suppression of the density effect in the vicinity of the upstream surface of the foil. The process of the density effect recovery inside the foil is defined by evolution of the TR field. This evolution comes down both to the radiation formation process and its attenuation inside the foil. In case the foil thickness exceeds both the radiation formation and attenuation lengths the density effect is fully manifested in K-shell ionization cross section at the downstream surface of the foil \cite{Spooner,Meyerhof}. In this case the cross section at the upstream surface (without the density effect impact) grows with the increase of the particle energy and exceeds the cross section at the downstream surface (with the density effect impact), which is energy independent. 

A much more diverse evolution of the electromagnetic field around a high-energy particle can take place in the target consisting of a large number of foils. In the present work we investigate the process of K-shell ionization by high-energy electrons in a periodic stack of parallel copper foils and CXR emitted in this case. It is shown that in such targets the K-shell ionization cross section, averaged over the target thickness, can be several times larger than the cross section at the upstream surface of a single foil where the density effect is absent. It is demonstrated that, generally, the average value of K-shell ionization cross section in multifoil targets is not defined just by the material of the foils and the electron energy, as might be expected, but depends on the target parameters, particularly, on separation between the foils. The corresponding CXR yield is therefore not unambiguously defined by the aggregated thickness of the target (total thickness of the foils), but depends on the number of foils in it and separation between them. It is also shown that, due to the mentioned increase of the ionization cross section and much smaller radiation attenuation, CXR yield from a multifoil target can considerably exceed the corresponding yield from a single foil of the same thickness as the aggregated thickness of the target.

It should be pointed out that in Ref.~\cite{Bak1986} the authors considered both theoretically and experimentally the influence of TR, generated by a high-energy electron in a stack of foils, on K-shell ionization cross section. Here the electrons crossed a stack of two thin foils, generating TR, and further hit (accompanied by the TR) a much thicker downstream foil. It is the ionization cross section inside the downstream foil which was studied in this case, while the stack of thin foils just played a role of TR radiator. In our work we consider a different statement of this problem, which corresponds to the case when K-shell ionization and CXR emission occur in the same multifoil target, where the TR process develops. Besides a series of new effects revealed for K-shell ionization cross section in this case (e. g., formation region effects, peculiar dependence on the number of foils in the target) it is also shown that presently the yield of CXR can be noticeably larger than in the scheme discussed in Ref.~\cite{Bak1986}, when TR and CXR are generated in different targets.

One of the differences of CXR from TR or coherent x-ray emission in crystals (i. e., parametric x-ray radiation \cite{Ter-Mikaelyan,Rullhusen,BaryshevskyBook} and diffracted transition radiation \cite{Caticha,ArtruNIMB}) is its homogeneous angular distribution (if neglect attenuation in the target). The latter types of emission by ultrarelativistic particles are highly concentrated respectively around the direction of the particle velocity and the Bragg direction. Thus, multifoil radiators based on the mechanisms of TR (see, e. g., Ref. \cite{Andronic} and references therein) and coherent x-ray emission in crystals \cite{PotylitsynPLA}, can provide a much higher radiation angular density than the corresponding CXR multifoil radiator. Nevertheless, the analysis made in the present work shows that even if the yield of CXR from a multifoil target is integrated over a rather small solid angle, it can be comparable to the yield provided by TR and coherent x-ray radiators.   

\section{Expressions for K-shell ionization cross section in a multifoil target}
\label{sec2} 

In order to calculate the average cross section of K-shell ionization by high-energy electrons incident upon a multifoil target we apply the approach developed in Refs. \cite{Bak1986,Sorensen}. In this approach the total cross section $\sigma_t$ is considered as a sum of two terms. The first one ($\sigma_c$) is associated with close collisions of the incident particles with atomic electrons, while the second one ($\sigma_d$) is due to distant collisions. The quantity $\sigma_c$ is not sensitive to the process of electromagnetic field evolution during the electron passage through the target. It can be calculated with the use of the well-known cross section of electron-electron scattering \cite{BerestetskiiQED}. The main attention in our work will be drawn to the quantity $\sigma_d$, which depends on the state of the field around the particle and varies along the particle path inside the target. At high electron energies this quantity can be calculated with the use of Weizs\"{a}cker-Williams method of equivalent photons \cite{Jackson,Ter-Mikaelyan} (the idea of the method was initially proposed by E. Fermi \cite{Fermi1924}). In this method the electromagnetic field around the electron (which consists of its proper Coulomb field, partially screened by polarization inside the foils, and the field of TR) is considered as an equivalent flux of photons moving together with the electron. In our case this flux consists of real photons of TR and virtual ones of the electron's proper field. In this method atomic ionization is described as photo effect which occurs as a result of interaction of these photons with the atomic electrons. 

In order to define the spectrum of equivalent photons it is necessary to calculate the Fourier component of electric field around the electron inside each foil of the target. Let the target consist of $N$ parallel foils of thickness $a$ separated by distance $b$ from each other. For simplicity, we will assume that the target is situated in vacuum, which occupies the region between the foils. Numerical estimations will be made for the case when the foils are made of copper. Let the electron move along the $z$-axis with the velocity $v$ and normally cross the foils. Let the upstream surface of the first foil along the electron path lie in the plane $z=0$. Inside the foils the Fourier component of the incident electron's proper field with frequency $\omega$  can be presented in the form: 
\begin{equation}
{\bf E}^{\textrm{pr}}_\omega({\bf r})=-\frac{ie}{\pi}\int d^2 q {\bf q} Q_f e^{i\omega z/v+i{\bf q}{\boldsymbol \rho}},
\label{Epr}
\end{equation}          
where 
\begin{equation}
Q_f=1/(q^2+\omega_p^2+\omega^2/\gamma^2),\nonumber
\label{Q_f}
\end{equation} 
$\gamma\gg1$ is the electron Lorentz factor, $\omega_p$ is the plasma frequency of the foils, $\boldsymbol \rho$ is the radius-vector of the observation point in the $xy$-plane, ${\bf r}=({\boldsymbol \rho},z)$. We use the system of units in which the speed of light $c=1$. The electron's proper field in vacuum has the same form as Eq. (\ref{Epr}) with the single substitution $Q_f\to Q_v$, where
\begin{equation}
Q_v=1/(q^2+\omega^2/\gamma^2). \nonumber
\label{Q_v}
\end{equation}
The field of TR generated upon the particle entrance into each foil (which propagates inside the foil) can be generally presented as follows:
\begin{equation}
{\bf E}^{tr}_\omega({\bf r})=\int d^2q{\bf E}'_\omega({\bf q})e^{i\omega z[1-(q^2+\omega_p^2)/2\omega^2]-\mu z/2+i{\bf q}{\boldsymbol{\rho}}},
\label{Etr}
\end{equation}  
where we took into account that $\omega_p\ll\omega$, as well as $q\ll\omega$ in the range of $q$, which make the main contribution to the integrals (\ref{Epr}) and (\ref{Etr}) at $\gamma\gg 1$. The quantity $\mu$ is the foil x-ray energy attenuation coefficient (attenuation of the field amplitude (\ref{Etr}) is described by the coefficient $\mu/2$). The field of TR generated upon the particle exit from the foil has the same form as Eq. (\ref{Etr}) with $\omega_p,\mu=0$. At high electron energies in the x-ray range of frequencies it is possible to neglect the TR emitted in the direction opposite to that of the electron velocity.

Consecutive application of boundary conditions for the electric field at vacuum-foil interfaces crossed by the electron leads, with the use of Eqs. (\ref{Epr}) and (\ref{Etr}), to the following expression for the Fourier component of the electric field around the electron inside the $n$-th foil:
\begin{equation}
\begin{aligned}
{\bf E}^{(n)}_\omega({\bf r})=&{\bf E}^{\textrm{pr}}_\omega({\bf r})+\frac{ie}{\pi}e^{i\omega(n-1)(a+b)/v}\int d^2q {\bf q} (Q_f-Q_v)\times \\
&F(q)e^{i\omega [z-(n-1)(a+b)][1-(q^2+\omega_p^2-i\mu\omega)/2\omega^2]+i{\bf q}{\boldsymbol \rho}}.
\label{Enth}
\end{aligned}
\end{equation}
Here 
\begin{equation}
F(q)=1-e^{-i\phi_v}(1-e^{-i\phi_f-\mu a/2})\frac{e^{-(n-1)[i(\phi_v+\phi_f)+\mu a/2]}-1}{e^{-[i(\phi_v+\phi_f)+\mu a/2]}-1}
\label{F}
\end{equation}  
with
\[
\phi_v=\frac{\omega b}{2}(\gamma^{-2}+q^2/\omega^2),~~~\phi_f=\frac{\omega a}{2}(\gamma^{-2}+q^2/\omega^2+\omega_p^2/\omega^2).
\]
The integrand in Eq. (\ref{Enth}) is analogous to the corresponding expression obtained in Ref. \cite{PotylitsynPLA} for the field of TR in a stack of thin crystalline foils (in the case of a periodic foil arrangement), but accounts for TR attenuation as well.

With the use of Eq. (\ref{Enth}) spectral density of the number of real and virtual photons, which cross the area $\rho_0<\rho<\infty$, is straightforwardly defined as
\begin{equation}
\frac{dN}{d\omega}=(4\pi^2\hbar\omega)^{-1}\int\limits_{\rho_0}^{\infty}|{\bf E}^{(n)}_\omega|^2 2\pi\rho d\rho.
\label{dNdw}
\end{equation}   
K-shell ionization cross-section due to distant collisions reads:
\begin{equation}
\sigma_d=\int\limits_{\omega_K}^{\infty}\frac{dN}{d\omega}\sigma^K_{ph}(\omega)d\omega,
\label{SigDist}
\end{equation}  
where $\hbar\omega_K$ is the minimum threshold energy required for the ionization, $\sigma^K_{ph}(\omega)$ is the cross section of K-shell photoionization and the value of $\rho_0$ is taken equal to $\sqrt{\hbar/2m\omega_K}$ \cite{Williams,Sorensen}, i. e. on the order of the Bohr radius of K-shell electron's orbit. 

The field (\ref{Enth}) is presented as an integral with respect to ${\bf q}$, which physically corresponds to the momentum transferred by this field to an atomic electron. Thus, in our case it is more convenient to make a restriction of integration interval in Eq. (\ref{dNdw}) not in the coordinate space, but in the momentum one (cf. Refs. \cite{ShulgaTrofymenkoPLA2012,TrofymenkoPLA2019}). This corresponds to integration with respect to $\rho$ on the interval $0<\rho<\infty$, but with respect to $q$ in the expression for ${\bf E}^{(n)}_\omega$ in Eq. (\ref{dNdw}) on the interval $0<q<q_0$ with $q_0=1/\rho_0$. Note that such a restriction concerns just the ${\bf E}^{\textrm{pr}}_\omega$ term in Eq. (\ref{Enth}), while the integrals containing the TR field in Eq. (\ref{dNdw}) are well convergent and do not require the analogous restriction. Substitution of Eq. (\ref{Enth}) into Eq. (\ref{dNdw}) leads to following form of the photon spectrum in the region around the incident electron inside the $n$-th foil:   
\begin{equation}
\begin{aligned}
\frac{dN}{d\omega}=&\frac{2\alpha}{\pi\omega}\bigg\{\ln\frac{q_0}{\sqrt{\omega^2/\gamma^2+\omega_p^2}}-1/2 \\
&+e^{-\mu[z-(n-1)(a+b)]}\int\limits_0^{\infty}dqq^3(Q_f-Q_v)^2|F(q)|^2 \\
&-2e^{-\mu[z-(n-1)(a+b)]/2}\int\limits_0^{\infty}dqq^3(Q_f-Q_v)Q_f \\
&\times\Re\bigg( e^{-i\omega[z-(n-1)(a+b)][\gamma^{-2}+(q^2+\omega_p^2)/\omega^2]/2}F(q)  \bigg)\bigg\},
\end{aligned}
\label{dNdwNonaveraged}
\end{equation} 
where $\alpha$ is the fine-structure constant. This expression presents the spectrum as a function of the distance $z$ along the electron path inside the target. Presently, the $n$-th foil occupies the region $(n-1)(a+b)<z<na+(n-1)b$. Eq. (\ref{dNdwNonaveraged}) shows that, generally, the photon flux around the electron varies from foil to foil. It also varies with the change of $z$ inside each foil. 

The terms from the first line in Eq. (\ref{dNdwNonaveraged}), being substituted to Eq. (\ref{SigDist}), give the cross section under the condition of the full value density effect, typical for a high-energy particle inside a thick foil on sufficiently large distance from its upstream surface \cite{Meyerhof,Spooner}. For instance, Eq. (\ref{dNdwNonaveraged}) reduces to its first line in the vicinity of the downstream surface of a single foil ($z\to a$, $N=n=1$) which thickness $a$ considerably exceeds the attenuation length of TR photons inside it at frequencies which contribute to Eq. (\ref{SigDist}). In this case it, naturally, coincides with the result predicted for this case in \cite{Sorensen} and rather nicely describes the available experimental data \cite{Spooner,Meyerhof}.

The second line in Eq. (\ref{dNdwNonaveraged}) defines the spectral density of TR photons in the region around the electron. The rest of the expression for $dN/d\omega$ originates from interference of the electron's proper field with the field of TR inside the foils. In the first foil of the target, in the vicinity of its upstream surface ($z\to 0$), $dN/d\omega$ reduces to the first line of Eq. (\ref{dNdwNonaveraged}), but with $\omega_p=0$. Being substituted to Eq. (\ref{SigDist}), such a photon spectrum, in accordance with Refs. \cite{Sorensen,Chechin}, results in the conventional K-shell ionization cross section unaffected by the density effect. As shown in \cite{Sorensen}, at high electron energies the value of this cross section, predicted by the currently applied method (which is the same as in \cite{Sorensen}), coincides with the experimental results (obtained in the measurements with very thin foils) with almost the same accuracy as the results of more rigorous theories.    

It is possible to make some simplification of Eq. (\ref{dNdwNonaveraged}), which considerably decreases the time of numerical calculations on the basis of this expression. Namely, at $q\gg\sqrt{2\omega/b}$ the value of $\phi_v$ is large and $F(q)$ becomes a rapidly oscillating function. Therefore, in this region it can be replaced by its average value. Averaging of $F(q)$ with respect to rapid oscillations of $e^{-i\phi_v}$ and $e^{-i(n-1)\phi_v}$, naturally, gives $\langle F\rangle=1$, while for $|F|^2$ after some calculations we obtain:
\begin{equation}
\begin{aligned}
\langle|F|^2\rangle=&1+4e^{-(n-1)\mu a/2}\bigg(\textrm{sh}^2\frac{\mu a}{4}+\sin^2\frac{\phi_f}{2}\bigg) \\
&\times\textrm{sh}\frac{(n-1)\mu a}{2}/\textrm{sh}\frac{\mu a}{2}.
\end{aligned}
\label{FavSq}
\end{equation}  
Note that in the above averaging procedure it was assumed that  $e^{-i\phi_f}$ is a slowly varying function compared to $e^{-i\phi_v}$ due to $a\ll b$, which usually takes place in practice. Besides, the analysis (see Sec. \ref{sec3}) shows that the most intense K-shell ionization in multifoil targets occurs when $a\sim2\pi\omega_K/\omega_p^2$ while the values of $\omega$ contributing to Eq. (\ref{SigDist}) are on the order of $\omega_K$. Contribution to the integrals in Eq. (\ref{dNdwNonaveraged}) is made by the values $q<\omega_p$. These imply that the typical values of $\phi_f$ in Eqs. (\ref{F}) and (\ref{FavSq}) are on the order of unity and it is not possible to make averaging over $e^{-i\phi_f}$ oscillations as well.

Thus, in numerical calculations on the basis of (\ref{dNdwNonaveraged}) for $q\gg\sqrt{2\omega/b}$ it is possible to use the corresponding averaged values $\langle F\rangle=1$ and $\langle|F|^2\rangle$ in the form (\ref{FavSq}) instead of $F$ and $|F|^2$. For smaller $q$ the non-averaged values of these quantities should be preserved. For $b\gg\omega/\omega_p^2$, which is usually the case in practice, the values $q\gg\sqrt{2\omega/b}$ occupy the major part of the effective integration region ($0<q<\omega_p$) and the above averaging allows noticeably shortening the calculation time. 

Let us introduce the quantity
\begin{equation}
l_v=2\gamma^2/\omega_K,
\end{equation}
which is the TR formation length in vacuum in the direction of the electron velocity taken at the minimal frequency $\omega=\omega_K$ which contributes to (\ref{SigDist}). If the spacing $b$ between the foils exceeds $l_v$, the region $q\lesssim\sqrt{2\omega/b}$, where the averaging of $F$ and $|F|^2$ is not possible, makes a very small contribution to the integrals in Eq. (\ref{dNdwNonaveraged}). In this case the averaged values of the discussed quantities can be applied in the whole range of integration with respect to $q$.

In order to compute the average ionization cross section in the multifoil target, it is necessary to average Eq. (\ref{dNdwNonaveraged}) with respect to $z$ inside each foil and further perform averaging of the obtained expression with respect to all the foils of the target (i. e., with respect to $n$). As a result, we obtain:     
\begin{align}
\frac{d\bar N}{d\omega}=&\frac{2\alpha}{\pi\omega}\bigg\{\ln\frac{q_0}{\sqrt{\omega^2/\gamma^2+\omega_p^2}}-1/2 \notag\\
&+\frac{1-e^{-\mu a}}{\mu a}\int\limits_0^{\infty}dqq^3(Q_f-Q_v)^2G(q) \notag\\
&-\frac{4}{a}\int\limits_0^{\infty}dqq^3(Q_f-Q_v)Q_f \notag\\
&\times\Re\bigg(\frac{1-e^{-a[\mu+i\omega(\gamma^{-2}+(q^2+\omega_p^2)/\omega^2)]/2}}{\mu+i\omega(\gamma^{-2}+(q^2+\omega_p^2)/\omega^2)}H(q)  \bigg)\bigg\}, 
\label{dNdwAv} 
\end{align}
where
\begin{equation}
\begin{aligned}
G(q)=&1+\frac{\textrm{sh}^2\frac{\mu a}{4}+\sin^2\frac{\phi_f}{2}}{\textrm{sh}^2\frac{\mu a}{4}+\sin^2\frac{\phi_f+\phi_v}{2}}\big[1+f(\mu a) \\
&-f(\mu a/2-i(\phi_f+\phi_v))-f(\mu a/2+i(\phi_f+\phi_v))\big] \\
&-2\Re(1-H(q)),
\end{aligned}
\label{G}
\end{equation}

\begin{equation}
\begin{aligned}
H(q)=&1-\frac{e^{-i\phi_v}(1-e^{-i\phi_f-\mu a/2})}{e^{-i(\phi_v+\phi_f)-\mu a/2}-1} \\
&\times\big[f\big(\mu a/2+i(\phi_f+\phi_v)\big)-1\big],
\end{aligned}
\label{H}
\end{equation}

\[
f(x)=\frac{e^{-Nx}-1}{N(e^{-x}-1)}.
\]   

Analogous averaging over quick oscillations of $e^{-i\phi_v}$, like in Eq. (\ref{dNdwNonaveraged}), can be performed in Eq. (\ref{dNdwAv}) as well. As a result, for the values $q\gg\sqrt{2\omega/b}$ (or for arbitrary $q$ in case $b>l_v$) the quantity $H(q)$ turns to unity, while $G(q)$ acquires the form:
\begin{equation}
\langle G(q)\rangle=1+2\frac{\textrm{sh}^2[\mu a/4]+\sin^2[\phi_f/2]}{\textrm{sh}[\mu a/2]}(1-f(\mu a)).
\label{GAv}
\end{equation}

Note that Eqs. (\ref{G}), (\ref{H}) and (\ref{GAv}) are valid for arbitrary $N$ and $a$. In case the aggregated target thickness $Na\gg \mu(\omega)^{-1}$ (this condition should be satisfied in the whole region of $\omega$ which contribute to Eq. (\ref{SigDist})) and $N\gg1$, all the functions $f$ can be neglected in these expressions. 

\section{Numerical estimation of the average cross section in a multifoil target}
\label{sec3}
In this section, we present the results of numerical estimation of K-shell ionization cross section in multifoil targets on the basis of the obtained expressions. We will consider here the average cross section $\bar\sigma_d$ in the target defined by Eqs. (\ref{SigDist}) and (\ref{dNdwAv}). As an example, the case of a target made of copper foils is investigated and the incident electron energy is taken equal to 5 GeV. Numerical values of $\sigma_{ph}^K$ we derive from the data on photon attenuation lengths $\mu^{-1}$ presented in Ref.~\cite{XrayWebSite} with the use of the relation $\sigma_{ph}^K=p_K\mu/n_a$ (presently, it is valid due to the fact that in the considered range of $\omega$ the attenuation is almost due to atomic photoelectric effect). Here $n_a$ is the atomic density of the foils and $p_K=(J_K-1)/J_K$ with $J_K\approx125/Z+3.5$ is the scaling factor \cite{Sorensen} defining the contribution of K-shell electrons to the total photoionization cross section of the atom ($Z$ is the atomic number of the considered element). For copper $\hbar\omega_K\approx 8.979$ keV. Due to a rather quick decrease of $\sigma_{ph}^K$ with the increase of $\omega$ it is the frequencies on the order of $\omega_K$ which make the main contribution to the integral in (\ref{SigDist}). For numerical estimations we will restrict the integration region here by $\hbar\omega_{\textrm{max}}=30$ keV, which allows taking into account almost all the ionization yield produced by the incident electron.  

\begin{figure}
	\includegraphics[width = 88mm]{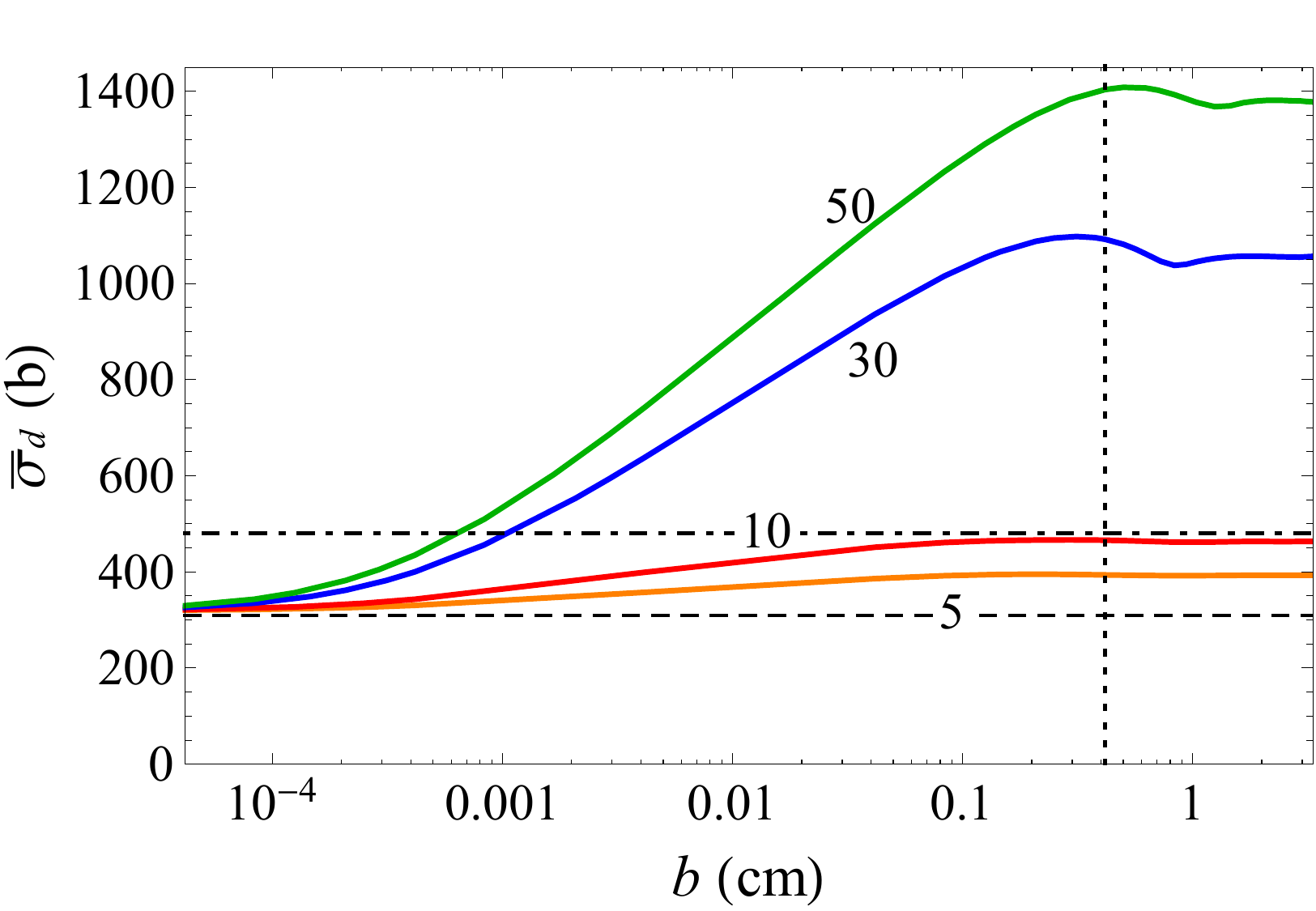}
	\caption{\label{fig1} Dependence of the average K-shell ionization cross section (due to distant collisions) in a multifoil copper target on spacing between the foils for different number of foils in the target (this number is indicated near each curve). The aggregated target thickness is $L\approx141~\mu\textrm{m}\approx1.33\mu^{-1}(\omega_{\textrm{max}})$, the incident electron energy is 5 GeV. Dashed line -- cross section under the condition of the full value density effect; dot-dashed line -- cross section in the absence of the density effect. Vertical line marks the value $b=l_v$.}
\end{figure}

It is illustrative to consider the average cross section $\bar\sigma_d$ in a multifoil target with a fixed aggregated thickness $L=aN$ (not taking into account the spacings between the foils) but varying other parameters, such as the number of foils $N$, which it consists of, and the spacing $b$ between them. Fig. \ref{fig1} shows the dependence of $\bar\sigma_d$ on spacing between the foils for four different foil numbers $N$ in the target. The spacing on the figure varies from about a micrometer up to several values of $l_v$. The target thickness $L$ equals $35\mu^{-1}(\omega_K)$, where the photon attenuation length is $\mu^{-1}(\omega_K)\approx4~\mu$m. Such value of $L$ is chosen to exceed the attenuation length $\mu^{-1}(\omega_{\textrm{max}})\approx106~\mu$m at the maximum frequency which is taken into account in the numerical estimation of (\ref{SigDist}). TR formation length inside the foils in the direction of the electron velocity at $\omega=\omega_K$
\begin{equation}
l_f=2\omega_K^{-1}/(\gamma^{-2}+\omega_p^2/\omega_K^2)
\end{equation}
is several times smaller than $\mu^{-1}(\omega_K)$ and equals about 1 $\mu$m. Thus, in case the target is solid and not split into foils, the average cross section inside it almost equals the one under the condition of the full value density effect (dashed line in the figure), associated with the first line in (\ref{dNdwAv}). The cross section in the thin boundary layer of such a target (for $z\ll l_f$), in accordance with Refs. \cite{Sorensen,Chechin}, demonstrates complete absence of the density effect (dot-dashed line in the figure).     

Generally, Fig. \ref{fig1} shows that the average cross section of K-shell ionization by high-energy electrons in a multifoil target is far from being defined by the material of the foils and the electron energy. It considerably depends on the number of foils which the target is split into. Moreover, at $b<l_v$ formation region effects take place for the ionization process \footnote{For a particular case of two foils, existence of such effects for the restricted ionization loss was pointed out in \cite{ShulgaTrofymenkoPLA2012}}, just like for the TR yield in multifoil radiators \cite{Cherry1974,TrofymenkoNIMB2020}. In this region the cross section logarithmically grows with the increase of $b$. We also see that for large $N$ (and, hence, small $a$), when the influence of TR upon $\bar\sigma_d$ is the most significant, the curves go through a small maximum before saturating to a constant value. Such a behavior is typical for the TR intensity in multifoil targets for $b\sim l_v$ \cite{GaribianBook}. It is a remnant of a much larger maximum in the dependence of TR spectral-angular density on $b$ due to constructive interference of the contributions of separate foils (for a more detailed discussion of the nature of this maximum see \cite{TrofymenkoNIMB2020}). Such a maximum almost (and sometimes completely) vanishes as a result of angle integration of this density (which corresponds to integration with respect to $q$ in our case). Let us also note that the value of $\bar\sigma_d$ for $N=5$ stays expectedly close to the cross section value under conditions of the full value density effect (dashed line) since in this case the thickness $a$ of separate foils still noticeably exceeds the TR attenuation length at $\omega=\omega_K$ (namely, $a\mu(\omega_K)=7$).

\begin{figure}
	\includegraphics[width = 88mm]{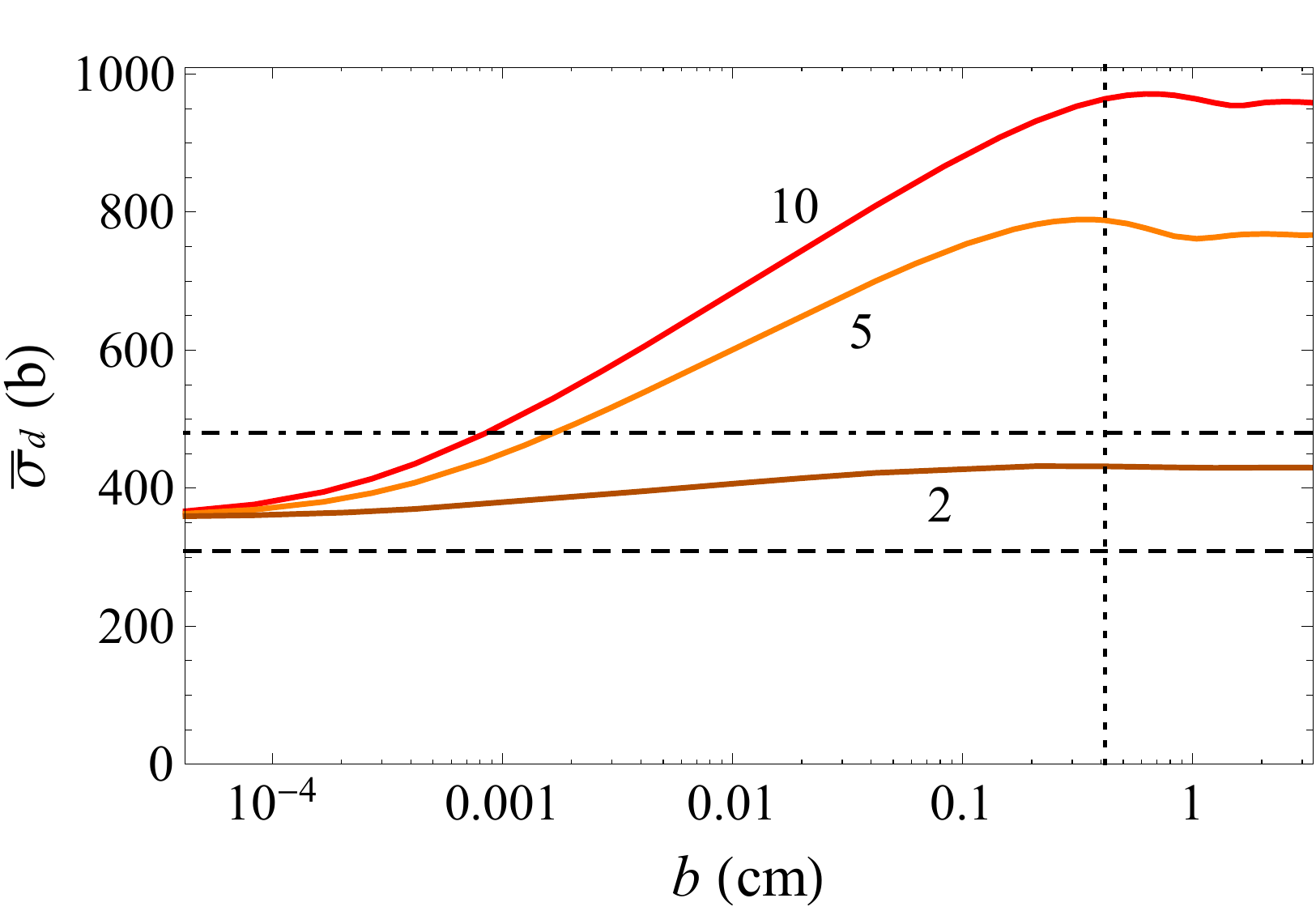}
	\caption{\label{fig1_1} The same as in Fig.~\ref{fig1}, but for $L=5\mu^{-1}(\omega_K)\approx20~\mu\textrm{m}$.}
\end{figure}

For comparison, in Fig. \ref{fig1_1} we also present a case in which the total thickness $L$ of the target is just several times larger than the attenuation length at the minimum frequency $\omega_K$ contributing to (\ref{SigDist}), namely, $L=5\mu^{-1}(\omega_K)\approx20~\mu$m. The results are shown for $N\leq10$, where the value of $a$ still exceeds several microns. 

One of the most interesting features of $\bar\sigma_d$, as shown in Fig. \ref{fig1} and Fig. \ref{fig1_1}, is that with the increase of the number of foils in the target (at fixed $L$) and spacing between them $\bar\sigma_d$ becomes larger than the cross section in the absence of the density effect. Such an effect takes place due to the fact that in the present case the photon attenuation length $\mu^{-1}$ noticeably exceeds the TR formation length $l_f$ inside the target (note that the difference between these lengths grows with the increase of $\omega$). It is mostly pronounced if the foil thickness $a$ is larger than $l_f$ and simultaneously $a\lesssim \mu^{-1}$ (as for $N=50$ in Fig. \ref{fig1}). In this case the field of TR generated upon the electron entrance into the first foil of the target becomes completely separated from the particle's proper field by the moment of the particle exit from this foil (since $a>l_f$) but is rather weakly absorbed (since $a\lesssim \mu^{-1}$). Thus, at the moment of its exit from the foil, the electron can again generate TR which is not suppressed by destructive interference with TR from the upstream surface (as would be if $a<l_f$). Hence, the electron impinges upon the second foil together with two almost full valued TR fields form each surface of the first foil (moreover, these fields can constructively interfere). Impinging upon the third foil, the particle is accompanied by four TR fields from the previous foils etc. The maximum (``stationary") number of foils $N_{\textrm{eff}}$ which contribute to the total TR field (which can be roughly considered as untouched by the absorption) accompanying the electron in the target is, naturally, estimated as $N_{\textrm{eff}}\sim(\mu a)^{-1}$. Such accumulation of TR photons (which are the source of K-shell ionization together with the electron's proper field) in the space around the electron results in the increase of ionization cross section.

The above reasoning can be most explicitly illustrated by Eq. (\ref{GAv}). In case TR attenuation can be neglected within the whole target ($\mu Na\ll 1$) it acquires the form:
\begin{equation}
\langle G(q)\rangle=1+2(N-1)\sin^2(\phi_f/2),
\label{GAverNoAbs}
\end{equation}    
which implies the linear growth of TR contribution (second line in Eq. (\ref{dNdwAv})) to $d\bar N/d\omega$ with the increase of $N$. (Presently, we consider $a$, and accordingly $\phi_f$, as constant and $L$ as increasing with $N$.) In a more realistic case the attenuation restricts such a linear growth at $N\sim N_{\textrm{eff}}$. Note that $N_{\textrm{eff}}$ can considerably vary with the change of $\omega$ within the interval which contributes to (\ref{SigDist}).  

\begin{figure}
	\includegraphics[width = 90mm]{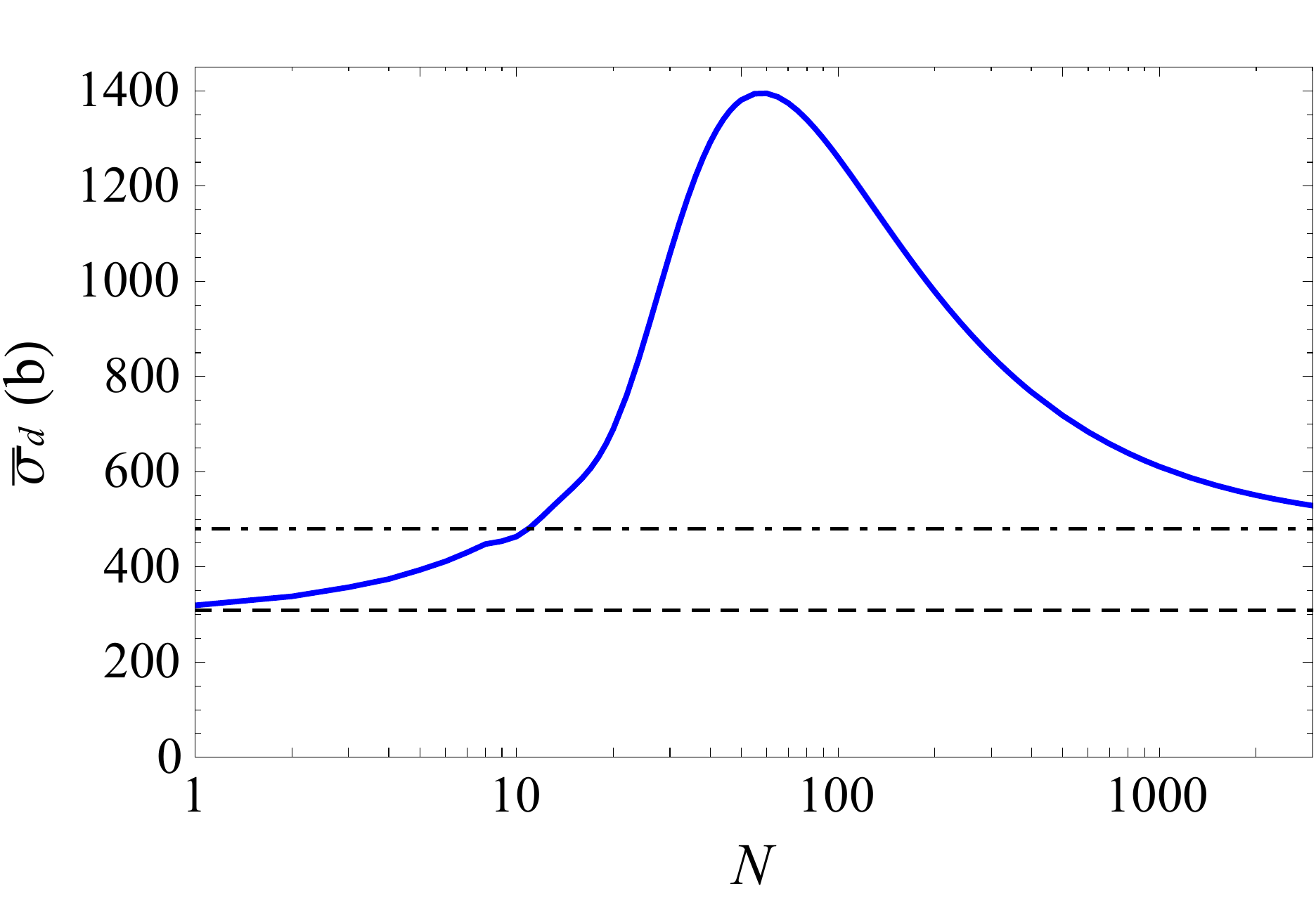}
	\caption{\label{fig2} Dependence of $\bar\sigma_d$ in a multifoil copper target of the fixed aggregated thickness $L\approx141~\mu$m on the number of foils in it for $b>l_F$. The incident electron energy is 5 GeV. Dashed and dot-dashed lines -- the same as in Fig. \ref{fig1}.}
\end{figure}

Fig.~\ref{fig2} shows the cross section dependence on the number of foils in the target for the fixed aggregated thickness of the latter. Here we consider the case $b>l_v$ when the cross section acquires its maximum value for each $N$. The presented here result is obtained with the use of the asymptotic form of Eq. (\ref{dNdwAv}), strictly valid for $b\gg l_v$, in which the averaged values of $G(q)$ and $H(q)$ are applied in the whole region of integration with respect to $q$. However, as Fig.~\ref{fig1} and Fig.~\ref{fig1_1} show, in the considered case $\bar\sigma_d$ becomes very close to its asymptotic value at $b\approx l_v$ or even less (the same holds for the cases with $N>50$, which are not presented in the mentioned figures). So, to a rather high accuracy, the discussed asymptotic form of Eq. (\ref{dNdwAv}) is presently valid beginning from $b\sim l_v$. For such $b$ the total length $L_t$ of the target, which includes spacings between the foils, is merely defined by these spacings and can be estimated as $L_t\approx(N-1)b$ (for $N\geq2$). For instance, we can choose $b=l_v$, which results in approximately 20 cm long target consisting of 50 foils (in this case, according to Fig.~\ref{fig2}, $\bar\sigma_d$ is close to its maximum value for the considered $L$).  

As Fig.~\ref{fig2} shows, for small $N$ and large $a$ the quantity $\bar\sigma_d$ is close to the cross section suppressed by the density effect, which is typical for thick foils. In the opposite case of large $N$ and small $a$ the quantity $\bar\sigma_d$ expectedly tends to the conventional cross section unaffected by the density effect, which is typical for ultrathin foils, as well as upstream surfaces of the foils of arbitrary thickness. The most interesting feature of the dependence presented in Fig. \ref{fig2} is that it is not monotonous and has a distinct maximum. The value of $\bar\sigma_d$ in this maximum is several times larger than the cross section in the absence of the density effect. The maximum corresponds to the foil thickness $a=L/N$ close to $\pi l_f(\omega_K)$, which is presently about $3~\mu$m. At such $a$ for $q\ll\omega_p$ we have $\phi_f=\pi$ (making the corresponding sine argument in (\ref{GAv}) and (\ref{GAverNoAbs}) equal $\pi/2$), which defines the condition of the most constructive interference between TR fields from the upstream and downstream surfaces of each foil.    

With the increase of electron energy the ratio of $\bar\sigma_d$ at the maximum to the cross section in the absence of the density effect monotonically increases (both these quantities increase with the energy as well) if we keep $b>l_F(\omega_K)$. Namely, it changes from 2.1 to 4 with the change of the electron energy from 1 GeV to 100 GeV ($L$ is the same as in Fig. \ref{fig1} and Fig. \ref{fig2}). At very high energies it should be, however, technically problematic to fulfill the condition $b>l_F$ due to large $l_F$ (at 100 GeV it exceeds 1.5 m). At such energies it might be inevitable to perform the measurements at $b\ll l_F$ where the above ratio is smaller due to formation region effects.       

\begin{figure}
	\includegraphics[width = 90mm]{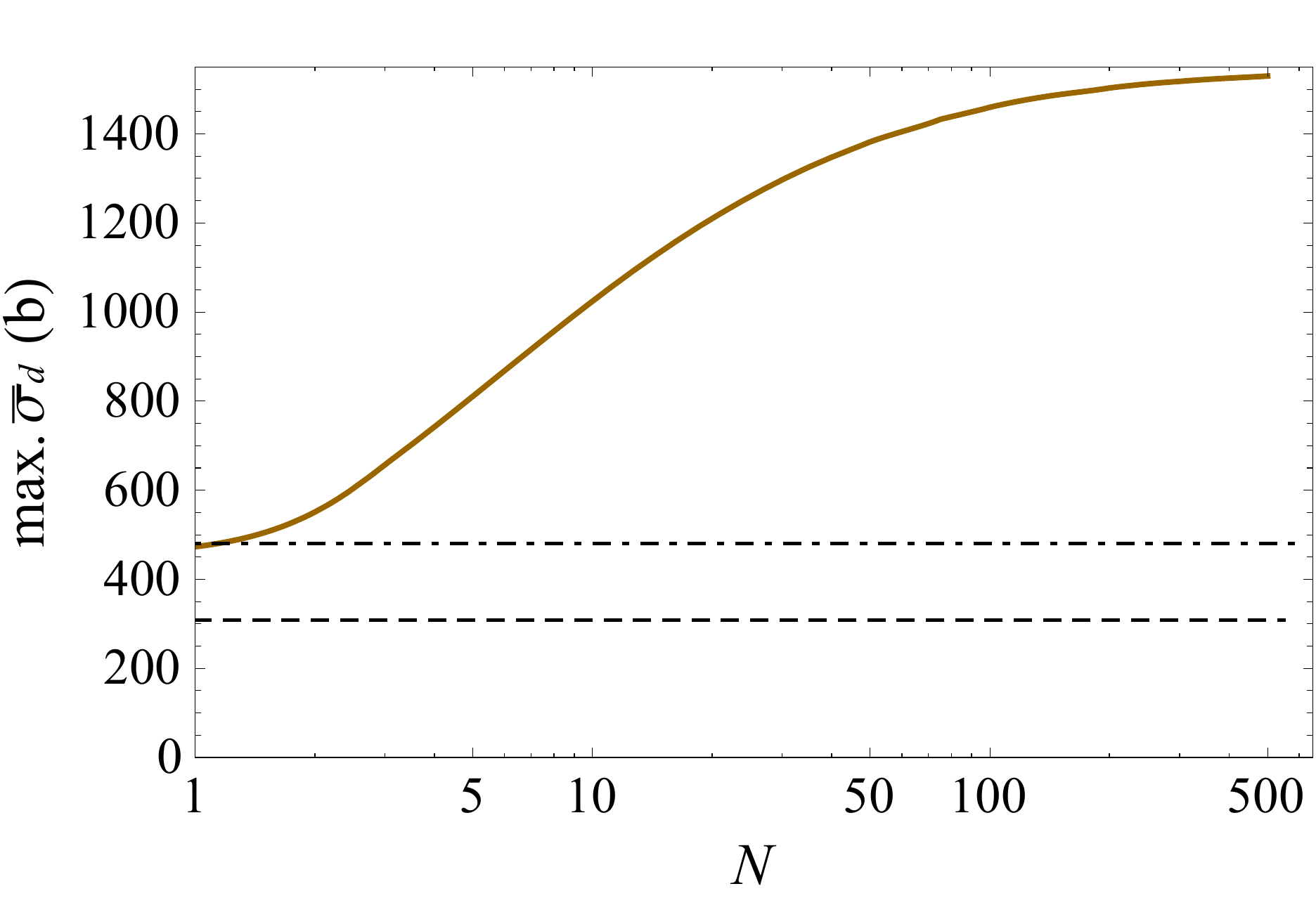}
	\caption{\label{fig3} Dependence of the maximum value of $\bar\sigma_d$ on the number of foils in the target. $L$ is not fixed and depends on $N$ and on the foil thickness $a$, which is chosen to be optimal. Spacing between the foils is $b>l_F$. The incident electron energy is 5 GeV. Dashed and dot-dashed lines -- the same as in Fig. \ref{fig1}.}
\end{figure}

Fig. \ref{fig3} demonstrates the dependence of the maximum value of $\bar\sigma_d$ on the number of foils in the target. Presently, the aggregated target thickness $L$ is not fixed and is defined by $N$ and the foil thickness $a$. For each $N$ the value of $a$ is chosen to maximize $\bar\sigma_d$. Such an optimal value of $a$ is almost independent on $N$ for $N\gg1$ and is slightly smaller than $2\pi\omega_K/\omega_p^2\approx3~\mu$m. Thus, in the considered case, the aggregated target thickness $L$ grows almost linearly with the increase of $N$. Fig. \ref{fig3} shows that $\bar\sigma_d$ does not saturate at small $N$ on the order of $(\mu(\omega_K) a)^{-1}\approx1.5$, as might be expected, but continues its monotonous increase at larger $N$. This happens due to the contribution to (\ref{SigDist}) from TR photons with $\omega>\omega_K$, which have larger attenuation lengths than $\mu(\omega_K)^{-1}$. For $N\gg1$ the discussed increase of $\bar\sigma_d$ is, however, rather slow and can be observed just for rather large variations of $N$ (e. g., $\bar\sigma_d$ changes by about 43\% with the increase of $N$ from 10 to 100).  

Finally, the contribution of close collisions $\sigma_c$ to the total cross section $\sigma_t$ can be estimated exactly in the same manner as it was done in \cite{Bak1986,Sorensen}: 
\begin{equation}
\sigma_c=\int\limits_{\omega_K}^{\infty}d\omega d\sigma_M/d\omega
\label{SigClose}
\end{equation}  
with $d\sigma_M/d\omega$ being the M{\o}ller cross section \cite{BerestetskiiQED} for the incident electron energy transfer to an atomic one (it contains additional factor of 2 accounting for the number of electrons at the K shell). This contribution does not depend on the electron energy (for $\gamma\gg 1$) and is not affected by the target polarization being the same throughout the electron path inside the foils. In the considered case it amounts to about 52 b.   

\section{CXR in a multifoil target}
\label{sec4}

The number of CXR photons emitted at the electron penetration through a multifoil target can be estimated on the basis of Eqs. (\ref{SigDist}) and (\ref{SigClose}) with $dN/d\omega$ in the form of Eq. (\ref{dNdwNonaveraged}). Angular density of the number of photons emitted from the whole target reads: 
\begin{equation}
\frac{dN_{\textrm{CXR}}}{do}=\frac{n_a w_f}{4\pi}\sum\limits_{n=1}^N\int\limits_0^a d\xi \sigma^{(n)}_t(\xi) g^{(n)}(\xi,\vartheta),
\label{Ncxr}
\end{equation} 
where $w_f$ is the K-shell fluorescence yield, defining the probability of photon emission as a result of vacancy filling at the ionized shell. For copper $w_f\approx0.44$. Also $\xi=z-(n-1)(a+b)$, while $\sigma^{(n)}_t(\xi)$ equals $\sigma_t(z)$ inside the $n$th foil. The function $g^{(n)}(\xi,\vartheta)$ accounts for CXR attenuation in the $n$th foil, where it is emitted, and in all the foils which it penetrates on its way to the detector. By $\vartheta$ we denote the angle between the direction of observation and the direction opposite to the one of the electron velocity. Let the foils have a circular shape with the radius $R$. In case the emission is considered in the backward hemisphere relative to the electron velocity ($\vartheta<\pi/2$), $g^{(n)}(\xi,\vartheta)$ has the following form:
\begin{equation}
\begin{aligned}
g^{(n)}(\xi,\vartheta)=&\exp(-\mu \xi/\cos\vartheta)\exp\big(-\textrm{Int}[R/b\textrm{tg}\vartheta]\mu a/\cos\vartheta\big)\\
&\times\eta\big(n-1-\textrm{Int}[R/b\textrm{tg}\vartheta]+\epsilon\big)\\
&+\exp(-\mu \xi/\cos\vartheta)\exp\big(-(n-1)\mu a/\cos\vartheta\big)\\
&\times\eta\big(\textrm{Int}[R/b\textrm{tg}\vartheta]-n+\epsilon\big).
\end{aligned}
\label{g}
\end{equation}    
Here $\eta(x)$ is the Heaviside step function and $\epsilon$ is an arbitrary number in the region (0,1) which ensures that $\eta(x+\epsilon)$ equals unity for $x=0$. The operator $\textrm{Int}[x]$ takes the integer part of $x$. In case the radiation is considered in the forward hemisphere ($\vartheta>\pi/2$), Eq. (\ref{g}) is still valid provided the following substitutions are made in it: $\xi\to a-\xi$, $n\to N-n+1$ and $\vartheta\to\pi-\vartheta$. 

CXR as a result of K-shell recombination in copper targets can be approximately considered as consisting of two monochromatic lines $K_\alpha$ and $K_\beta$. Due to relatively small difference between the frequencies of these lines, as well as due to a relatively small contribution of $K_\beta$ line to the total CXR yield in copper (about 0.14 from that of $K_\alpha$ \cite{Holtzer}), we will take the attenuation coefficient $\mu(\omega)$ in (\ref{g}) at the frequency of $K_\alpha$ line $\hbar\omega_{K\alpha}\approx 8.05$ keV. In fact, it is the frequency of $K_{\alpha1}$ line dominating in the doublet of $K_{\alpha1}$ and $K_{\alpha2}$ lines, which we presently consider as indistinguishable. For CXR attenuation length in this case we have $\mu^{-1}\approx22~\mu$m.  
\begin{figure}
	\includegraphics[width = 90mm]{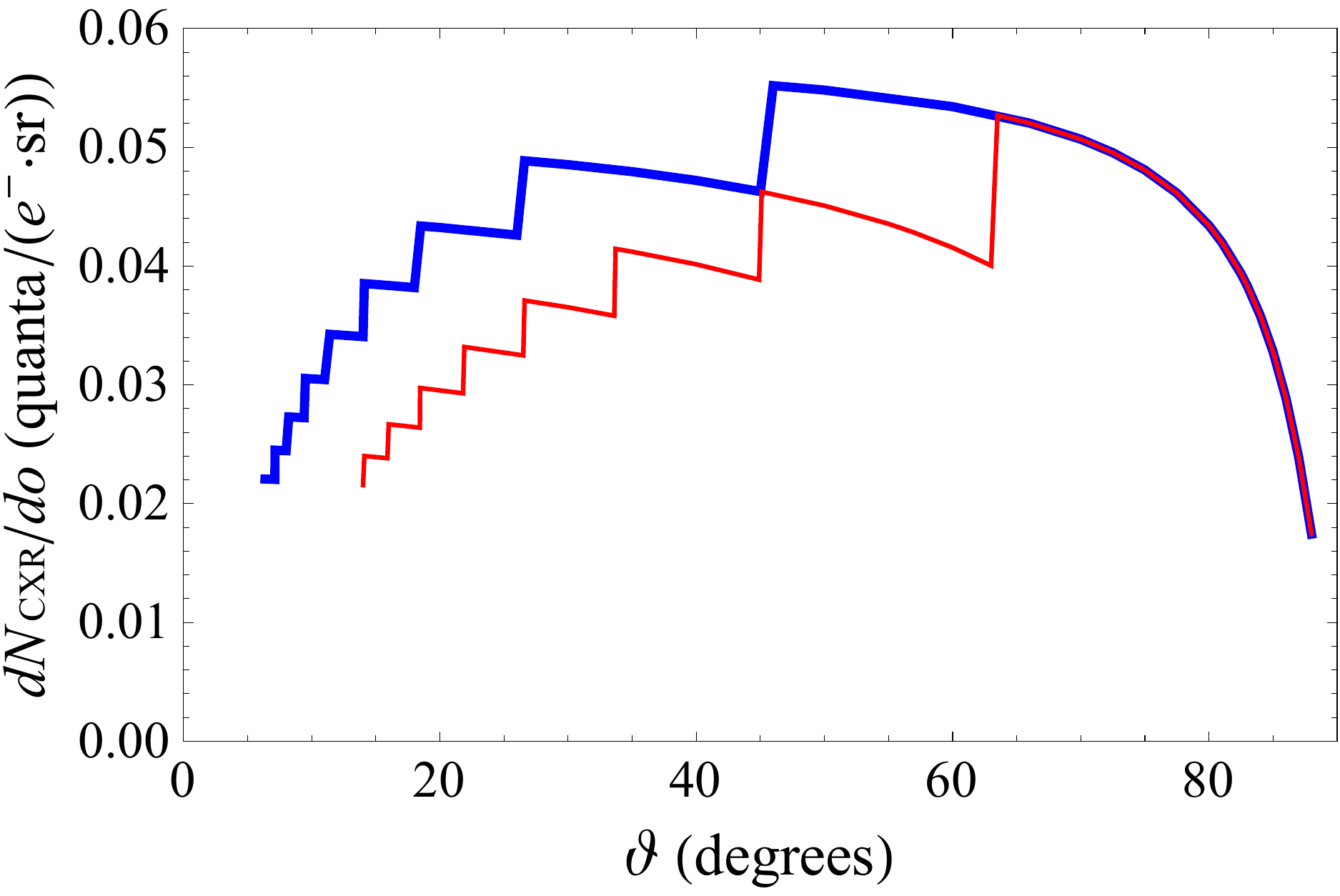}
	\caption{\label{fig4} Angular distribution of CXR from the multifoil copper target with $N=50$, $a\approx 2.8~\mu$m and $R=0.5$ cm. Thick blue line -- $b=0.5$ cm, thin red line -- $b=0.25$ cm. The incident electron energy is 5 GeV.}
\end{figure} 

Fig. \ref{fig4} demonstrates the angular distribution of CXR generated by 5 GeV electrons in the multifoil target of aggregated thickness $L=35\mu^{-1}(\omega_K)\approx 141~\mu$m under the condition when $\bar\sigma_d$ is close to its maximum value ($a\approx2\pi\omega_K/\omega_p^2$ and $b\sim l_v$). The difference of CXR angular distribution from the spherically symmetric one, described by $g^{(n)}(\xi,\vartheta)$, is due to radiation attenuation in the foils which it needs to penetrate in order to escape from the target. Each jump of the emission intensity, taking place at $\vartheta=\arctan(R/kb)$, where $k$ is a positive integer, is caused by a new foil which appears on the way of the photons with the decrease of $\vartheta$. 

With the increase of the number of foils in the target, provided $a$ remains close to its optimal value of $2\pi\omega_K/\omega_p^2$ and $b\sim l_v$, for the major part of the values of $\vartheta$ (excluding the ones close to 0 and $\pi$) the CXR angular density $dN_{\textrm{CXR}}/do$ grows roughly in proportion to $N\bar\sigma_d$ (the increase of $\bar\sigma_d$ in this case is presented in Fig.~\ref{fig3}). This happens due to the fact that for such $\vartheta$ the radiation has to cross a rather small number of foils in order to escape the target and the increase of $N$ above this number does not increase the influence exerted by the absorption inside the foils on this radiation.  

According to Fig. \ref{fig4}, at $b=0.5$ cm the maximum number of photons is emitted in the direction of $\vartheta\approx50^\circ$. If consider electron normal incidence upon a single foil of the same thickness as the whole target ($a\approx 141~\mu$m), or any other thickness satisfying the condition $a\gg\mu^{-1}(\omega_{K\alpha})$, for the same angle $\vartheta$ one obtains the photon density $dN_{\textrm{CXR}}^{(1)}/do\approx 0.0017~\textrm{quanta}/(e^{-}\cdot \textrm{sr})$. This result can be obtained with the use of general expressions (\ref{SigDist}), (\ref{dNdwNonaveraged}), (\ref{SigClose}) and (\ref{Ncxr}) with $n=N=1$. Thus, we see that due to the discussed above effect of $\bar\sigma_d$ increase (Fig. \ref{fig2}), as well as due to smaller radiation attenuation, it is possible to obtain a considerable enhancement of CXR yield in a multifoil target compared to such a yield in a single foil of the same aggregated thickness. Particularly, under the considered conditions $(dN_{\textrm{CXR}}/do)/(dN_{\textrm{CXR}}^{(1)}/do)\approx32$.

\section{Comparison with other types of emission}
\label{sec5}

It is also illustrative to compare the yield of CXR in the discussed case with yields of some other types of x-ray emission in multifoil targets. In this section we make the comparison of CXR yield under conditions corresponding to Fig. \ref{fig4} (the case of $b=0.5$ cm and $\vartheta$ in the vicinity of $50^\circ$) with the yields of TR and, in case the foils are crystalline, coherent x-ray emission. In all cases we consider the targets containing the same number of foils ($N=50$) as in the discussed case of CXR, i. e. making estimations for the same target size. The separation $b$ between the foils is assumed to exceed the formation length $l_F$, which under considered conditions is on the order of several millimeters, in order to avoid radiation suppression due to formation region effects. Though, with the decrease of $b$ the angle-integrated yields of the considered types of emission decrease rather slowly, following the logarithmic dependence, as in Fig. \ref{fig1}. Thus, the decrease of $b$ down to the values several times smaller than $l_F$ does not considerably change the yield.   

\subsection{Transition radiation}
\label{subsec5A}

As a typical example of TR radiator we consider a stack of 50 thin aluminium foils. The radiation spectral density $dN_{\textrm{TR}}/d\omega$ in this case is defined by the well-known expressions \cite{Cherry1974,Artru1975}, which we apply in the form presented in \cite{TrofymenkoNIMB2020} (formula (1)). The thickness of the foils is chosen equal to 10 $\mu$m, which corresponds to the highest TR intensity at the maximum of its spectrum situated at $\hbar\omega\approx 12$ keV. The electron energy is 5 GeV, as before. Since CXR is a monochromatic emission, it is more practical to compare it not with the ``pure'' TR from the radiator, which has a broad spectrum, but with the monochromatic radiation which can be obtained on the basis of this radiator. In practice such monochromatization of TR is often accomplished by its further Bragg diffraction on a single crystal. We will presently consider diffraction of TR by (111) planes of a silicon crystal oriented in such way that the Bragg condition is satisfied for the photon energy in the vicinity of 12 keV corresponding to the maximum of the TR spectrum. For this, the crystal surface should to be inclined at the angle $\theta_B\approx 9.5^\circ$ (Bragg angle) relative to the electron velocity (the plane (111) is assumed to be parallel to the surface). In this case the yield of the diffracted TR can be estimated as \cite{Pinsker,Chaikovska}
\begin{equation}
\begin{aligned}
N_{\textrm{DTR}}&=\int d\omega R(\omega)dN_{\textrm{TR}}/d\omega \\ &=\frac{16}{3}\bigg( \frac{\omega_B}{cg} \bigg)^2|\chi_{\bf g}|P\omega_B\bigg(\frac{dN_{\textrm{TR}}}{d\omega}\bigg)_{\omega=\omega_B},
\end{aligned}
\label{Ndtr}
\end{equation} 
where the speed of light $c$ is preserved for convenience. Here $R(\omega)$ is the crystal reflection coefficient and integration is performed over the narrow frequency region around the Bragg frequency $\omega_B\approx cg/(2\sin\theta_B)$, corresponding to the so-called Darwin table. Here and further we neglect small variation of $\omega_B$ with the observation angle in the angular region which comprises the major part of the emitted photons. Presently, $g$ is the absolute value of reciprocal lattice vector ${\bf g}$ of the considered set of planes, $\chi_{\bf g}$ are the coefficients in the Fourier series expansion of the crystal dielectric susceptibility with respect to ${\bf g}$, $P$ is the polarization factor which under the considered condition ($\theta_B\ll1$) can be set equal unity. For the chosen set of crystallographic planes of the silicon crystal the coefficient in front of $\omega_B dN_{\textrm{TR}}/d\omega$ in (\ref{Ndtr}) equals $1.77\cdot10^{-4}$ and for the total yield of diffracted TR we obtain $N_{\textrm{DTR}}\approx1.5\cdot10^{-4}~\textrm{quanta}/e^{-}$. The angle integration of the TR incident upon the crystal was performed over the region of angles (0, $\omega_p/\omega_B$) relative to the electron velocity, which comprises almost all the emitted photons (presently, $\omega_p$ is the plasma frequency of aluminium foils)\footnote{For completeness, let us also note that the total number of TR photons emitted by the radiator in the region from 5 to 40 keV, comprising the major part of the photon yield, approximately equals unity, which is a typical yield for the considered type of radiators.}. Thus, the estimated number of photons is emitted within the cone with the opening angle of $2\omega_p/\omega_B\approx0.3^\circ$. 

CXR is not such a narrowly directed emission like TR and has a much lower angular density. Nevertheless, in a multifoil target it is possible to obtain photon yields of CXR, comparable to that of TR radiators (after the considered monochromatization of this emission), even within rather small solid angles. For instance, under the conditions, typical for Fig. \ref{fig4}, at $\vartheta\approx50^\circ$ the same CXR yield of $1.5\cdot10^{-4}~\textrm{quanta}/e^{-}$, as the estimated above $N_{\textrm{DTR}}$, corresponds to the opening angle of about $3^\circ$. This provides a possibility to apply CXR by high-energy particles in multifoil targets as a source of monochromatic x-ray photos, like TR. The advantage of such a source is associated with broad angular distribution of the emission, which allows catching the photons in the wide range of directions. 

Another way to obtain monochromatic emission on the basis of a TR multifoil radiator is to let the TR (together with the electrons which generate it) fall on a downstream foil and produce CXR there. It corresponds to the approach applied in \cite{Bak1986} where the authors studied CXR from a 25 $\mu$m copper foil produced by electrons which preliminarily crossed two thin (3.26 $\mu$m) upstream copper foils generating TR. Though, for a large number of foils in the TR-radiating target this approach provides a noticeably smaller CXR yield than the one considered in the present work (when TR and CXR are generated in the same multifoil target). It is due to the fact that in the TR radiator with a large number of foils, due to absorption, just a part of these foils near the downstream end of the radiator contribute to the emission of TR which escapes the radiator and falls on the downstream foil. In the case when CXR is generated in the multifoil target, all the foils contribute to the radiation yield. For instance, let us consider the same target as in Sec.~\ref{sec4} (50 copper foils of 2.8 $\mu$m thickness separated by the distance of 0.5 cm from each other) as a TR radiator and estimate the angular density of CXR emitted when the TR, generated in it by 5 GeV electrons, falls on a copper foil of thickness $a\gg\mu^{-1}(\omega_{K\alpha})$. Substituting the corresponding TR spectrum to Eq. (\ref{SigDist}), taking into account its absorption inside the downstream foil, and applying Eq. (\ref{Ncxr}) with $N=1$ (presently, we should put $\sigma_d=\sigma_t$), for $\vartheta=50^\circ$ we get $dN_{\textrm{CXR}}^{(1)}/do\approx 0.0046~\textrm{quanta}/(e^{-}\cdot \textrm{sr})$, where we also took into account the contribution from the electron itself, calculated at the end of Sec.~\ref{sec4}. This result is more than ten times smaller than the one presented in Fig.~\ref{fig4} for the same $\vartheta$. Let us also note that for the parameters of the two-foil TR radiator and downstream target considered in Ref.~\cite{Bak1986}, the result of our calculation of the average K-shell ionization cross section in the target quite nicely coincides with the corresponding results (theoretical and experimental) presented in the mentioned work (this cross section is about 8 \% higher than the conventional value without the influence of the density effect).         

\subsection{Coherent x-ray emission. Ultrathin foils}
\label{subsec5B}

Let us now compare the yield of CXR in a multifoil target with the yield of coherent x-ray emission in a stack of ultrathin crystalline foils, discussed in \cite{PotylitsynPLA}. The latter type of emission can be considered as a result of Bragg diffraction of the field around the charged particle moving in crystal on crystallographic planes. The results of diffraction from separate planes coherently add up to each other and form an almost monochromatic and narrowly directed pulse of x-ray emission in the vicinity of the Bragg direction. Generally, this radiation can be divided into two parts associated with diffraction of virtual photons of the particle's proper field (parametric x-ray radiation or PXR \cite{Ter-Mikaelyan,Rullhusen,BaryshevskyBook}) and of real TR photons emitted at the particle entrance into the crystal (diffracted TR \cite{Caticha,ArtruNIMB}). In sufficiently thin crystals these contributions are inseparable and interfere with each other. In an isolated crystal, which thickness is smaller than both the x-ray extinction length $l_{\textrm{ext}}$ and TR formation length $l_f$ inside it, the radiation spectral-angular distribution $d^2N_{\textrm{coh}}/d\omega do$ resembles that of PXR without the account of medium polarization influence on it \cite{Nasonov,TrofymenkoPhysRev2018}. If the crystal thickness is not much smaller than $l_{\textrm{ext}}$ the emission can be considered as approximately monochromatic for the fixed observation direction (like in thicker crystals) and $d^2N_{\textrm{coh}}/d\omega do$ can be easily integrated with respect to frequency to obtain:
\begin{equation}
\frac{dN_{\textrm{coh}}}{do}\approx\frac{\alpha}{4\pi}\frac{a\omega_B}{c}|\chi_{\bf g}|^2\frac{\vartheta^2}{(\vartheta^2+\gamma^{-2})^2\sin^2\theta_B},
\label{Ncoh}
\end{equation}       
where $\vartheta$ is the angle between the direction of observation and the Bragg direction, $a$ is the crystalline foil thickness. Note that here we neglected the dependence of $dN_{\textrm{coh}}/do$ on the radiation polarization, which is possible either for $\theta_B\ll1$ or $\pi-2\theta_B\ll1$. 

As an example of radiator we will presently consider a stack of 50 parallel silicon foils of thickness $a=0.5~\mu$m. As in Subsec. \ref{subsec5A}, we consider the radiation produced on the set of (111) planes, parallel to the surfaces of the foils. If choose the foil inclination angle $\theta_B$ equal to $14.3^\circ$, the emitted radiation frequency will be $\omega_B\approx8$ keV, just like in the case of CXR from copper foils. The coherent x-ray emission yield in this case can be calculated via integration of the expression for radiation spectral-angular distribution derived in \cite{PotylitsynPLA} (generally, radiation attenuation should be taken into account here as well). We will, however, apply a simpler approach, which allows to approximately estimate the yield in the considered case. Presently, $l_f\approx3~\mu$m and the fact that it is noticeably larger than $a$ allows neglecting the TR generated as a result of the electron passage through the foils (due to strong destructive interference of contributions from the upstream and downstream surfaces of each foil). It is also worth noting that in this case such TR is not accumulated from a large number of foils, as in the previously considered case of CXR. This is due to the fact that presently the extinction length, which plays the role of attenuation length for TR photons which undergo diffraction, is just $l_{\textrm{ext}}\approx1.6~\mu$m \cite{XrayWebSiteExtLength}. (Note that $l_{\textrm{ext}}$ describes the exponential decrease of the TR field strength due to diffraction and should be divided by 2 in order to describe the analogous decrease of the TR intensity.) Since $l_{\textrm{ext}}<l_f$, TR is diffracted before any noticeable amount of it is emitted by a series of neighbouring foils. As a result, the emission yield in each foil is approximately the same as in an ultrathin isolated crystal and is defined by (\ref{Ncoh}). The total number of photons emitted from the whole target inside the cone with the opening angle $2\vartheta_{\textrm{max}}=6\omega_p/\omega_B\approx1.3^\circ$ is around $2\cdot10^{-5}~\textrm{quanta}/e^{-}$. Presently, the value of $\vartheta_{\textrm{max}}$ is chosen to be larger than the typical value $\vartheta=\omega_p/\omega_B$ corresponding to the maximum of PXR angular distribution. The maximum of the expression (\ref{Ncoh}) at $\vartheta=\gamma^{-1}$ is much closer to the axis of the above cone. Under the conditions typical for Fig. \ref{fig4} the same number of CXR photons is emitted inside the cone with the same opening angle of about $1.3^\circ$.    

\subsection{Coherent x-ray emission. Thick foils}
\label{subsec5C}

The yield of coherent x-ray emission from the target with the same number of foils naturally increases if take much thicker foils. The maximum yield in this case is achieved if the foil thickness in the direction of the electron motion $a/\sin\theta_B$ exceeds not only $l_{\textrm{ext}}$ and $l_f$, but the attenuation length $\mu^{-1}(\omega_B)$ of the emitted photons inside the foil as well. In this case the emission consists of well-separated contributions of PXR (with the maximum at $\vartheta=\omega_p/\omega_B$) and diffracted TR (with the maximum at $\vartheta=\gamma^{-1}$). PXR contribution from each foil is defined by (\ref{Ncoh}) with the substitutions $a\to \mu^{-1}(\omega_B)$ and $\gamma^{-2}\to\gamma^{-2}+\omega_p^2/\omega_B^2$. For arbitrary separation between the foils the contribution of diffracted TR from each foil is defined by the expressions derived in Ref. \cite{TrofymenkoPRAB2019}. If $b>l_v$ it is just the doubled value of the diffracted TR yield from an isolated thick foil, defined by the conventional formula \cite{Caticha,ArtruNIMB,Chaikovska}. The doubling of the yield from the $n$th foil occurs due to additional contribution from the TR emitted at the electron exit from the $(n-1)$th foil, which is diffracted in the $n$th foil. The situation is, certainly, different for the very first foil of the target, but for $N\gg1$ this fact can be neglected. The total yield of PXR and DTR inside the cone with the same opening angle as before ($6\omega_p/\omega_B\approx1.3^\circ$) in this case amounts to about $5\cdot10^{-4}~\textrm{quanta}/e^{-}$ ($10^{-5}$ photons for each foil). The same number of CXR photons is emitted inside the cone with the opening angle of about $6^\circ$.      

\section{Conclusions}
\label{Conclusions}

In the present work we considered the process of K-shell ionization by high-energy electrons in multifoil copper targets and characteristic x-ray radiation emitted in this case. It is shown that in such targets the average K-shell ionization cross section $\bar\sigma_d$ is influenced by formation region effects analogous to the ones typical for transition radiation in multifoil targets. In case the separation $b$ between the foils is smaller than the TR formation length $l_v$ in the region between the foils, such effects lead to the logarithmic increase of $\bar\sigma_d$ with the increase of $b$. The rate of this increase depends on the number of foils $N$, which the target consists of, provided the aggregated thickness $L$ of the target is fixed. Due to peculiarities of evolution of the electromagnetic field around the high-energy electron during its motion through the target, the value of $\bar\sigma_d$ can become several times larger than the conventional K-shell ionization cross section without the density effect impact. Dependence of $\bar\sigma_d$ on the number of foils in the target is studied both for the cases of fixed and variable aggregated target thickness. The optimal target parameters, which correspond to the maximum value of $\bar\sigma_d$, are obtained. The angular density of CXR emitted in this case is calculated taking into account its attenuation in the foils of the target. It is shown that in the considered case the radiation is much more intense than in the case of electron incidence on a single foil of the same aggregated thickness. The yield of CXR in the considered case is compared to the yields of some other types of x-ray emission in multifoil targets, which have much larger angular density. This includes transition radiation as well as coherent x-ray emission in a stack of equally oriented crystalline foils. It is shown that, due to the effect of $\bar\sigma_d$ increase and small radiation attenuation, the number of CXR photons emitted from a multifoil target within a rather small solid angle can be comparable to the photon yield typical for the mentioned types of emission, which enables application of CXR in the considered scheme as an x-ray photon source. The advantage of the source of this kind is associated with the broad radiation angular distribution. Due to this fact, the photons can be caught in the wide range of directions and the corresponding installation does not require a rigorous alignment. It is also shown that the considered scheme (when CXR is emitted from the multifoil target itself) can provide a much higher photon yield than the scheme (which particular case was studied in \cite{Bak1986}) in which the multifoil target is applied as an upstream TR radiator, while CXR is emitted from the downstream foil.


\begin{acknowledgments}

The work was partially supported by Projects No. C-2/50-2020 and No. F30-2020 of the National Academy of Sciences of Ukraine (budget program ``Support for the Development of Priority Areas of Scientific Research'', 6541230). 

\end{acknowledgments}

\bibliography{apssamp}

\end{document}